# Low-Threshold Surface-Emitting Whispering-Gallery Mode Microlasers

Andrey Babichev, Ivan Makhov, Natalia Kryzhanovskaya, Sergey Troshkov, Yuriy Zadiranov, Yulia Salii, Marina Kulagina, Mikhail Bobrov, Alexey Vasil'ev, Sergey Blokhin, Nikolay Maleev, Leonid Karachinsky, Innokenty Novikov, and Anton Egorov

*Abstract—* **We report on microlasers based on high-quality micropillars with lasing on whispering-gallery modes. Usage of low-absorbing $Al_{0.2}Ga_{0.8}As/Al_{0.9}Ga_{0.1}As$ distributed Bragg reflectors as well as the smooth pillar sidewalls allows us to realize whispering-gallery modes lasing by excitation and collection of emission in the pillar axis direction. As a result, simultaneous whispering gallery modes lasing (comb-like structure) in the wavelength range of 930–970 nm is observed for 3–7 μm pillar diameters. Increase the temperature up to 130 K results single-mode lasing for 5 μm pillars with about 8000 cold cavity quality-factor and 240 μW estimated threshold excitation power.**

## I. INTRODUCTION

The semiconductor micropillars (with micron-sized diameters) with microcavity formed by distributed Bragg reflectors (DBRs) and quantum dots (QDs) active region could demonstrate the lasing on the vertical (light propagates across the active region) direction via axial modes. Whispering-gallery modes (WGM) propagating along the active region that are generally observed in semiconductor microdisks [1–9] or ring cavities [10–12], could be also supported in high-quality micropillars [13]. In fact, WGM mode is weakly confined in vertical direction by mirrors and strongly confined in horizontal direction by micropillar sidewalls. If one can realize the high quality (Q)-factor for WGM modes [14, 15] the ratio of high Q-factor and small WGM-mode volume [15] (due to WGM mode location close

Manuscript received XX XXXX 2024; revised XX XXXX 2024; accepted XX XXXX 2024. Date of publication XX XXXX 2024; date of current version XX XXXX 2024. The authors from Ioffe Institute acknowledge support in part by the grant of the Russian Science Foundation No. 22-19-00221, https://rscf.ru/project/22-19-00221/ for the structure design, MBE epitaxy, and the study of photoluminescence and lasing spectra. N. Kryzhanovskaya thanks the Basic Research Program at the HSE University for support of the study of photoluminescence spectrum measured at 90° tilted sample of microcavity structure. *(Corresponding authors: Andrey Babichev.)*

Andrey Babichev, Sergey Troshkov, Yuriy Zadiranov, Yulia Salii, Marina Kulagina, Mikhail Bobrov, Alexey Vasil'ev, Sergey Blokhin, and Nikolay Maleev are with Ioffe Institute, 194021 Saint Petersburg, Russia (e-mail: a.babichev@mail.ioffe.ru).

Ivan Makhov, and Natalia Kryzhanovskaya are with HSE University, 190008 Saint Petersburg, Russia.

Leonid Karachinsky, Innokenty Novikov, and Anton Egorov are with ITMO University, 197101 Saint Petersburg, Russia.

Color versions of one or more of the figures in this article are available online at http://ieeexplore.ieee.org

Digital Object Identifier

to the sidewall surfaces [14]) is comparable with axial mode. As a result, quantum electrodynamics phenomena can be observed [14]. Moreover, as it previously has been mentioned [16] lasing mode energy is stable for excitation power as six times exceeding the threshold value. Additionally, WGM energies are very sensitive to pillar diameter [15]. Thus, the precise tuning of the lasing mode energy (emission wavelength) by diameter is possible. Low threshold excitation power allows one to suppose that the pillars with surface-emitting WGMs lasing can be used to form spectrally uniform (spectral homogeneity) micropillars array suitable for optical reservoir computing (RC) [17–20]. Previously, the dense micropillars arrays were fabricated with tuning the wavelength of fundamental vertical mode by refining diameter of the individual micropillar [19], but the same adjustment could be realized by changing the wavelength of WGM mode with diameter tuning.

To date, the WGM lasing in vertical microcavity was demonstrated for the case of GaAs/Al(Ga)As micropillars with In(Ga)As QDs-based active region at optical [14–16, 21–23] and electrical excitation scheme [24].

In general, the WGM lasing in micropillars was realized by excitation and collection of emission from the pillars sidewalls [14, 16]. The disadvantage of sidewalls excitation is that just low fraction of light absorbed by the gain medium (about few percent [16]).

The use of the excitation in the pillar axis direction with collection of emission from the pillar sidewalls (60° degrees) was also applied [21, 22]. The collection of emission under angle (non-axis direction) restricts the usage of WGMs micropillars for optical RC, where diffractively coupled (via optical injection locking) lasers array [19] is necessary. In fact, for optical RC the surface-emitting dense spectrally homogeneous low-threshold microlasers array is required [19].

Surface-emitting WGMs micropillars based on vertical microcavity were realized at 14 K temperature by excitation and collection of emission in the pillar axis direction [15], but the high threshold excitation power restricts its usage for RC. The excitation of the QDs just through the structure sidewalls (by Ni mask located on top DBR) resulted in the threshold excitation power about 100 mW [15].

Here, we conducted microphotoluminescence studies of micropillar cavities based on $Al_{0.2}Ga_{0.8}As/Al_{0.9}Ga_{0.1}As$ DBRs and InGaAs QDs using the excitation and collection of emission in the pillar axis direction. Due to the high mirror loss for the axial modes and the smooth sidewalls obtained with the optimized dry etching process, the low-threshold (as low as 180 µW optical excitation power) WGMs lasing in the 930–970 nm spectral range was demonstrated at 77 K. Moreover, single-mode WGMs lasing was observed with increase the temperature to 130 K.

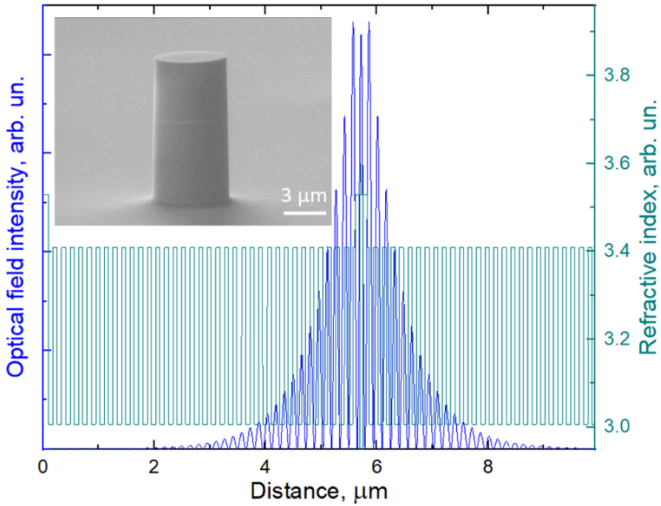

**Fig. 1.** Optical field intensity versus distance from the substrate. The inset demonstrates the scanning electron microscope image of micropillar.

## II. EXPERIMENTAL METHODS

The microcavity structure was grown by molecular-beam epitaxy on semi-insulated GaAs substrate. Bottom and top mirrors consist of 35 and 27 low-absorbing pairs of $\lambda/4$n-thick $Al_{0.2}Ga_{0.8}As/Al_{0.9}Ga_{0.1}As$ layers [18, 20] instead of general used GaAs/Al(Ga)As DBRs to realize the excitation and collection of emission in the pillar axis direction. The active region was based on three layers of self-assembled Stranski-Krastanow QDs formed from $In_{0.5}Ga_{0.5}As$ layers. The QD layers were separated by 20 nm thick GaAs barriers to eliminate vertical stacking. The QD based active region was embedded in the center of one-$\lambda$ thick (about 302 nm) cavity (cf. Figure 1). The cavity design was optimized aimed to achieve high exciton conversion efficiency (efficient excitation of the QD array through the GaAs matrix) as well as minimize the threshold pump power (the GaAs absorbing layer thickness is about 210 nm). The micropillars were formed by photolithography and dry etching down in depth about 9 µm (cf. Figure 1, inset). The optimization of photoresist mask and dry etching process resulted in a smooth sidewall with vertical mesa etching profile (less than 1 degree deviation from a normal).

The sample was mounted on a copper holder in a Montana closed cycle optical cryostat with the possibility of changing the sample temperature in the range of 4–300 K. As it was mentioned previously [18], the optical absorption edge of $Al_{0.2}Ga_{0.9}As$ at 77 K temperature is about 1.75 eV (about 707 nm wavelength). Thus, aimed increasing the power conversion efficiency it is preferable to use the optical excitation at the wavelength above this value. As a result, to measure the emission spectra an 808 nm semiconductor laser diode, operating in a continuous-wave mode was used. Laser emission was focused onto the sample using a system of mirrors and its optical power was varied by a rotary gradient filter. Focusing of the laser beam on the sample, as well as the collection of photoluminescence emission, was carried out using a Mitutoyo MPlan Apo NIR microobjective (50x, NA 0.42). Focused spot size was 7 µm aimed to realize the same excitation power density for all studied pillars diameters. Moreover, usage of 7 µm spot diameter allows one to suppress the vertical mode that is generally excited at 1–2 µm spot size [18]. As result, all excitation powers given in Section III were recalculated on real pillar diameter with assumption that there is no absorption on the pillar sidewalls. In fact, even at sidewalls excitation scheme the fraction of light absorbed by the gain medium is about few percent [16].

To collect the emission, the Andor Shamrock 500i grating monochromator equipped with thermoelectrically cooled silicon DU 401A BVF back Illuminated CCD matrix was used. Usage of a threaded diffraction grating 1200 lines per mm limits the spectral resolution about 0.05 nm (about 68 µeV at 950 nm line) of used monochromator with 500 mm focal length [18].

## III. EXPERIMENTAL RESULTS

Figure 2 demonstrates the photoluminescence spectra measured for pillars with different diameters. With increase of the excitation power one can see the saturation of the integrated intensity for standard axial modes of the vertical micropillar cavity located between 970–985 nm for all studied pillar diameters that proofs that there is no lasing via axial mode. Actually, the vertical mode is located close to the pillar center in opposite to WGM modes that is observed in experiment. In fact, the lasing is observed close to 897 nm for 2 µm pillar. Increase the pillar diameter up to 3 µm reveals to simultaneous lasing at 929 nm and 951 nm. This multi-mode lasing via WGM modes is related to the inhomogeneous broadening of the QD-based active region [14, 21]. In fact, the FWHM of photoluminescence spectrum measured at 90° tilted sample of microcavity structure was about 56 meV. The 4 µm pillar demonstrates lasing close to 918 nm, and 951 nm. Lasing lines at 951 nm and 964 nm were observed for 5 µm pillar. A comb-like structure of the emission spectra with lasing lines located at 936 nm, 948 nm, 960 nm, and 972 nm was observed for 6 µm pillar diameter. The multi-mode lasing via WGM modes was demonstrated for 7 µm pillar with the lasing at 947 nm, 957 nm, and 967 nm. The separation between WGMs is slightly changed with emission energy from 13.46 meV to 13.59 meV (cf. Figure 3, inset). Previously, the same behavior was demonstrated [14, 21] and was related with energy ($E$) dispersion of effective refractive index $n_{eff}$ [15, 21]. The free spectral range, FSR of the low-energy side of the lasing spectra coincides to inversely proportional diameter dependence (cf. Figure 3) roughly expressed by equation [14, 21]: FSR ~ $hc/\pi Dn_{eff}$, where $h$ – Plank constant, $c$ – the speed of light, $D$ – pillar diameter. In opposite to results [15, 18, 25–27], the birefringence in spectra (split WGMs) was not observed for all studied pillars diameter. Previously, the birefringence in spectra was related to the pillar ellipticity or to scattering on localized defects on the pillar circumference which lifts the ±m two-fold degeneracy [15]. In fact, the unintentional ellipticity for all studied pillars is no more than 0.3 % for 3–7 μm diameter range determined by the scanning electron microscope study.

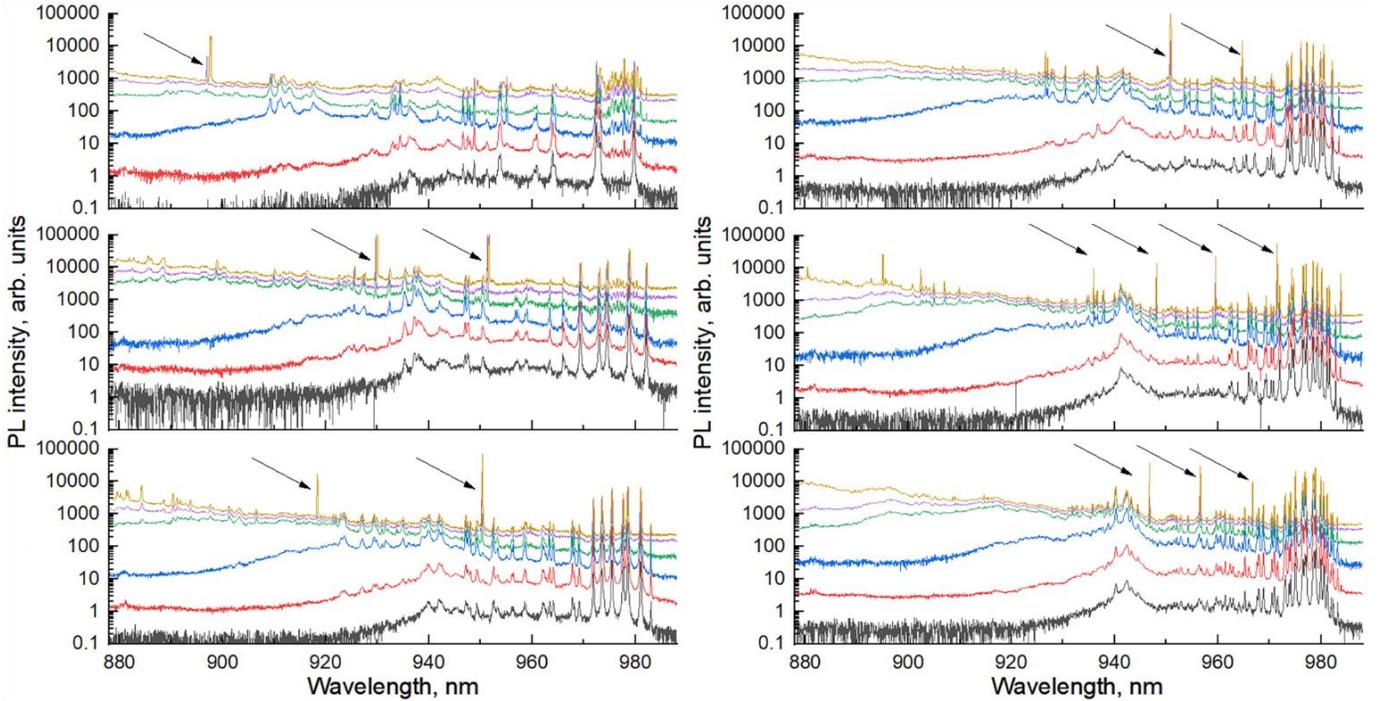

**Fig. 2.** PL spectra (semi logarithmic scale) measured at 77 K temperature. Left part demonstrates results for 2 μm (top panel), 3 μm (middle panel) and 4 μm (bottom panel). Right part demonstrates results for 5 μm (top panel), 6 μm (middle panel) and 7 μm (bottom panel). The spectra are measured at excitation power 4.3 μW, 43 μW, 0.43 mW, 2.1 mW, 4.5 mW and 11.6 mW, respectively for 7 μm pillar. The excitation power for 6 μm, 5 μm, 4 μm, 3 μm and 2 μm pillars can be recalculated by multiply the same values given for 7 μm pillar on 0.73, 0.51, 0.33, 0.18 and 0.08 coefficient, respectively. Arrows demonstrate the lasing lines.

The Gaussian line shape fitting was applied to define the power characteristics as well as Q-factor [19]. Saturation of background PL as well as superlinear dependence of integrated PL intensity of WGM modes versus excitation power were observed for all studied pillars diameters. Typical clear S-shaped power characteristics (in double-log scale) are presented on Figure 4 for the case of 2 μm pillar (897 nm line) and 5 μm pillar (951 nm line) measured at 77 K. Moreover, clear S-shaped power characteristic along with reduced linewidth to 77 μeV signaturing the laser transition for 5 μm pillar even at 130 K temperature (cf. Figure 5).

The spontaneous emission factor (betta factor, β) can be determined by the solution of laser-rate equation [18, 28]: $P_{ext.} = A\gamma/\beta[BP_{out}(1+\xi)(1+\beta BP_{out})-\beta\xi BP_{out}]$, with γ as the cavity decay rate, B as the scale factor between the intracavity photon number and the output power $P_{out}$, A as the scaling factor between the pump rate and the excitation power $P_{ext}$. The ξ value was evaluated by expression $\xi = n_0\beta/\gamma\tau_{sp}$, where $n_0$ is the number of excitons at transparency and $\tau_{sp}$ is the spontaneous exciton lifetime [29]. As it was mentioned previously [20], the cavity decay rate can be expressed by equation: $\gamma = 2\pi\nu_L/Q$ with $\nu_L$ as the lasing frequency, which along with ξ value determined as in

Ref. 21 results a β-factor of about 1.2% for 5 μm pillar measured at 77 K. The estimated value of threshold excitation power determined as $P_{th} = A\gamma/2\beta[\xi(1-\beta)+1+\beta]$ is about 420 μW for 5 μm pillar measured at 77 K. It makes sense to estimate the threshold value by the analyze of the full Voigt profile of emission lineshape with excitation power [30]. The Lorentzian and Gaussian components of linewidth as well as the shape factor (μ factor) can be determined [30]. It was observed that the μ factor is close to 1 at excitation power up to 460 μW that reveals about clear Lorentzian lineshape below threshold for 5 μm pillar. At 560 μW excitation power of the Gaussian component becomes comparable with the Lorentzian one. As a result, the cold (empty) cavity Q-factor, determined as emission energy divided by the linewidth at the lasing threshold [15, 21], is around 10000 for 5 μm pillar at 77 K. Increase of the excitation power to 2.8 mW leads to 13400 Q-factor due to the onset of the QD gain [15]. The estimated value of β-factor for 2 μm pillar is about 1.1 % measured at 77 K. As a result, the evaluated threshold excitation power is about 180 μW for 2 μm pillar that coincides to empty Q-factor of about 5000. By the same approach one can estimate the threshold excitation power of about 240 μW for 5 μm pillar measured at 130 K that coincides to empty Q-factor less than 7800. To clarify decrease of laser threshold with temperature we are measured the gain-to-cavity (GCD) detuning on microcavity structure. The negative GCD (~9 nm) was evaluated at 77 K. In opposite, about 3 nm positive GCD value was obtained at 130 K.

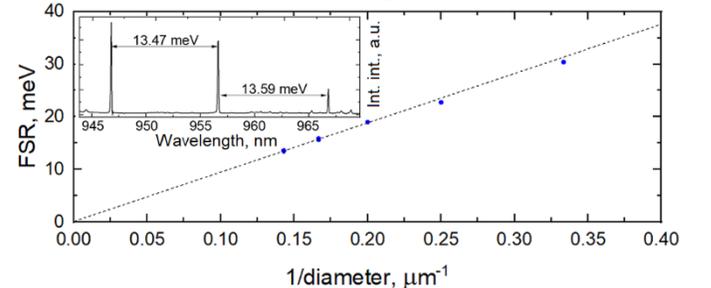

**Fig. 3.** Free spectral range versus inverse pillar diameter. The inset demonstrates enlarged spectrum for 7 μm pillar measured at 11.6 mW excitation power.

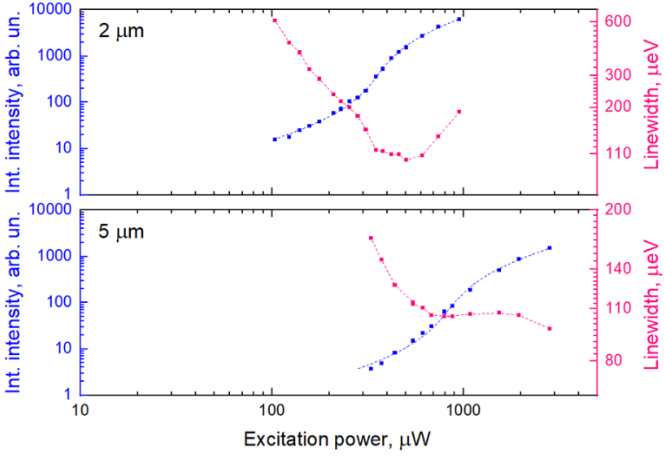

**Fig. 4.** The input-output measurement measured at 77 K for 2 μm pillar (top panel, left Y axis) and 5 μm pillar (bottom panel, left Y axis). The spectral linewidth (the Gaussian fit) values are presented on right Y axis.

The spectral linewidth determined at 77 K is shown on Figure 4. One can see an abrupt narrowing of the emission line with increase of the excitation power, followed by saturation and even growth (in the case of 2 μm pillar). A narrowing of linewidth resulted to the onset of stimulated emission. At high pumping levels, the linewidth could be influenced by both the nonlinearity of the active region gain and heating effects. In fact, different linewidth versus excitation power behavior was demonstrated for 2 and 5 μm pillars (cf. Figure 4). The 2 μm pillars demonstrate a narrowing of the spectral linewidth with further changing to inhomogeneous broadening of the laser linewidth due to thermal effects [18]. In fact, excitation-power-dependent mode slight red shift (~ 250 μeV) for 2 μm pillar is demonstrated with increase the excitation power up to 490 μW. Further increase the excitation power to 930 μW reveals to drastically red shift of mode position (from -250 μeV to -1500 μeV) due to unwanted heating. In opposite, the 5 μm pillar has not demonstrated the inhomogeneous broadening of the laser linewidth at high excitation power. In fact, the maximal excitation power density was about 14 kW/cm² which is approximately two times lower than for 2 μm pillar case (~30 kW/cm²). The absence of linewidth broadening at high excitation power is also demonstrated at 130 K temperature (cf. Figure 5). As it was mentioned previously [18], the usage of low-absorbing top DBR instead of general used GaAs/Al(Ga)As pairs allows one to minimize the undesired heating due to optical absorption in top DBR. Aimed to clarify this issue we have analyzed the excitation-power-dependent mode shift, presented on Figure 6. At 77 K temperature increase of the excitation power up to about 1.1 mW results in 65 μeV blueshift of the mode position that can be related to band filling and the plasma effect as it was previously supposed [18, 24]. Further increase of the excitation power to 2.8 mW reveals the slight redshift of the mode position (cf. Figure 6) that is caused by thermal effect dominating over the plasma effect. To summarize, increase the excitation power density from curve roll-over value (~6 kW/cm²) up to 30 kW/cm² results in a very stable emission with just about 31 μeV redshift at excitation powers up to 7 times above the laser threshold. The close redshift of the mode position (–26 μeV, from 55 μeV to 29 μeV) was demonstrated previously for 5.4 μm pillar lasing on fundamental vertical mode [31] in excitation power density range from 6 kW/cm² to 30 kW/cm².

The opposite behavior of the mode position was demonstrated at 130 K temperature (cf. Figure 6). The absence of the mode position shift is demonstrated up to 0.5 mW results that plasma effect compensates the thermal effect. Further increase the excitation power reveals that thermal effect starts to dominate. At the same time the absolute red shift is not exceeds 170 μeV at excitation powers up to 12 times above the laser threshold (from 1.2 kW/cm² to 14.1 kW/cm²). The thermal mode shift can be estimated by the shift of mode position versus cryostat temperature [16]. In fact, increase the 5 μm pillar temperature from 77 to 130 K reveals to about 3.4 meV red shift (~ -65 μeV/K). As a result, the estimated rise of the pillar temperature is about 3 K up to 12 times the laser threshold.

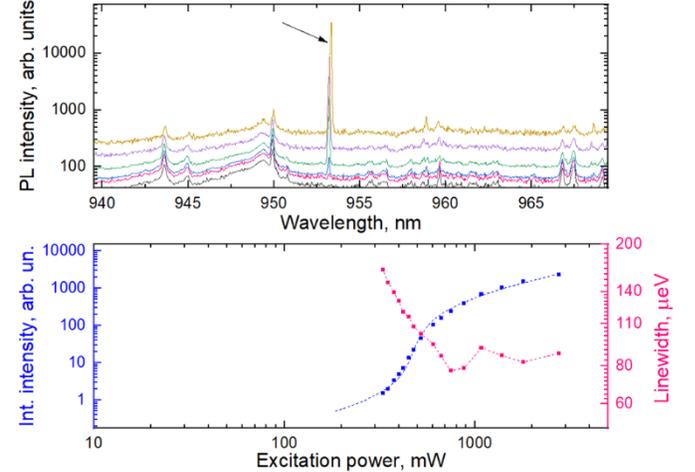

**Fig. 5.** PL spectra (semi logarithmic scale) for 5 μm pillar measured at 130 K temperature (top panel). The spectra are measured at excitation power 0.22 mW, 0.33 mW, 0.42 mW, 0.61 mW, 1.1 mW and 2.8 mW, respectively. Arrow demonstrates the lasing line (close to 953 nm); The input-output curve (953 nm line) measured at 130 K for 5 μm pillar (bottom panel, left Y axis). The spectral linewidth (the Gaussian fit) values are presented on right Y axis.

## IV. CONCLUSION

We report on realization of surface-emitting WGMs lasing with low threshold excitation power in GaAs microcavities. The clear *S*-shape input-output characteristic was observed up to 130 K thanks to the smooth pillar sidewalls and with low-absorbing modified DBRs.

Previously, the WGMs lasing in pillar microcavities was just demonstrated at cryogenic temperatures (at 4 K–14 K [14, 15, 16, 21, 24]. Moreover, surface-emitting WGMs lasing was realized just through the structure sidewalls resulted in the threshold excitation power about 100 mW [15]. Herein, the surface-emitting lasing with excitation through top mirror was realized with 240 μW threshold excitation power at 130 K. This value is about 6 times less than the previously reported value of the threshold excitation power for micropillars with the lasing on fundamental vertical mode ($P_{th}$=1.5 mW as 130 K [20]). The small shift of mode energy (~170 μeV at 130 K) was demonstrated with increase the excitation power about twelve times.

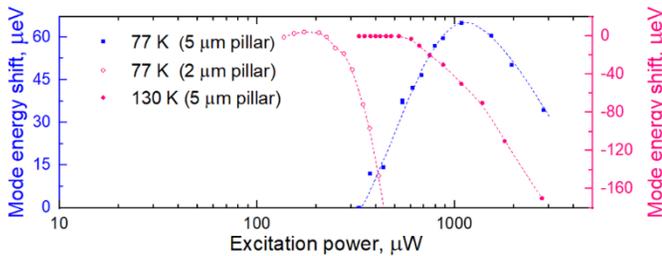

**Fig. 6.** Mode energy shift as a function of excitation power for 2 μm and 5 μm pillars.

## Acknowledgment

The low temperature studies were partially carried out on the equipment of the large-scale research facilities "Complex optoelectronic stand".

**Andrey Babichev** received the Ph.D. degree in condensed matter physics from the Ioffe Institute, St. Petersburg, Russia, in 2014. He has authored or co-authored 182 papers published in refereed journals and conference proceedings. His research interests include semiconductor heterostructures and lasers based on them. Repeated DAAD scholarship as well as Metchnikov scholarship holder. He received the Academia Europaea awards (the Russian Prizes) in 2016. He also admitted the Ioffe Prize in 2020 and 2023.

**Ivan Makhov** was born in Saint Petersburg, Russia, in 1992. He received the master's and Candidate of Science (Ph.D.) degrees from Peter the Great St. Petersburg Polytechnic University (SPBPU), Saint Petersburg, in 2015 and 2021, respectively. He is currently a Researcher at the International Laboratory of Quantum Optoelectronics, National Research University Higher School of Economics, Saint Petersburg. His research interests include the studies of optical and electrooptical phenomena in semiconductor-based nanostructures in infrared and terahertz spectral ranges.

**Natalia Kryzhanovskaya** received the Engineering degree from Saint Petersburg State Electrical Engineering University, Saint Petersburg, Russia, in 2002, the Candidate of Science (Ph.D.) degree from the Ioffe Institute in 2005, and the D.Sc. degree from Saint Petersburg Academic University, Saint Petersburg, in 2018. She is currently the Head of the Laboratory of Quantum Optoelectronics, National Research University Higher School of Economics, Saint Petersburg. Her research interests include the study of passive and active optoelectronic devices based on nanostructures.

**Sergey Troshkov** interests include scanning electron microscope analysis of semiconductor heterostructures.

**Yury Zadiranov** interests include semiconductor heterostructures and its processing technology.

**Yulia Salii** interests include semiconductor heterostructures and its processing technology.

**Marina Kulagina** was born in Zlatoust, Russia, in 1960. She received the Graduate degree from the Chemical-Techological Department, Leningrad Forest Academy, St. Petersburg, Russia, in 1982. She started working at A.F.Ioffe Physico-Technical Institute, Russian Academy of Science, St. Petersburg. She is currently a Research Fellow at the Centre of Nanoheterostructure Physiscs, Ioffe Physical-Technical Institute, St. Petersburg. She has coauthored more than 50 papers. Her research interests include research and development of postgrowth technologies, including photolithography, thin-film deposition, and etching of different semiconductor devices. Mrs. Kulagina received the Russian Academy of Science Diploma for the successful work.

**Mikhail Bobrov** received the Diploma (M.S. equivalent) degree in physics from Peter the Great St. Petersburg State Polytechnic University, St. Petersburg, in 2014. He has co-authored 74 papers published in refereed journals and conference proceedings. His research interests include the theoretical modeling and characterization of the different optoelectronic devices, including vertical-cavity surface emitting lasers, photodiodes, and single photon sources.

**Alexey Vasil'ev** interests include semiconductor heterostructures and its growth technology.

**Sergey Blokhin** received the Ph.D. degree in semiconductor physics from the Ioffe Institute, St. Petersburg, in 2006. He has authored or co-authored over 180 papers published in refereed journals and conference proceedings. His current research interests include technology and characterization of III–V semiconductor nanostructures and development of the different optoelectronic devices, including vertical-cavity surface emitting lasers, photodiodes and single photon sources.

**Nikolay Maleev** received the Ph.D. degree in Engineering from the St. Petersburg State Electrotechnical University "LETI", St. Petersburg, in 1997. He has authored or co-authored over 160 papers and 7 patents. His current research interests include design, modeling and technology of III–V heterostructure devices, including vertical-cavity surface emitting lasers and quantum dot single photon sources.

**Leonid Karachinsky** received the Ph.D. degree in semiconductor physics from the Ioffe Institute, St. Petersburg, Russia, in 2004, and Dr. Sci. degree in Engineering from the ITMO University, St. Petersburg, Russia, 2021. Since 2021 he became a Professor and Deputy Director at the Institute of Advanced Data Transfer Systems, ITMO University. He has co-authored over 100 papers. His research interests include semiconductor heterostructures and lasers based on them.

**Innokenty Novikov** received the Ph.D. degree from the Ioffe Institute, Russian Academy of Sciences, St. Petersburg, in 2005. He has co-authored over 50 papers. His current research interests include development and investigation of the different types of semiconductor lasers and theoretical modeling of optical properties of semiconductor lasers.

**Anton Egorov** received the Diploma degree from the Electrical Engineering Institute in Leningrad in 1987 and the Ph.D. and Dr. Sci. degrees in 1996 and 2011, respectively. He is a Professor at ITMO University, St. Petersburg, Russia. He has authored over 410 papers published in refereed journals and conference proceedings. His area of experience includes molecular beam epitaxy; III–V semiconductor heterostructures; GaAs- and InP-based (In,Ga,Al) (As,N) quantum well and quantum dots heterostructures; edge- and vertical-cavity surface emitting lasers (infrared); and III–V heterostructures for microelectronics.